\documentclass[nojss]{jss}

\usepackage[utf8]{inputenc}
\usepackage{xspace}
\usepackage{appendix}
\usepackage{breakurl}

\newcommand{\R}{\textsf{R}\xspace}
\newcommand{\Java}{\textsf{Java}\xspace}
\newcommand{\C}{\textsf{C}\xspace}
\newcommand{\Cpp}{\textsf{C++}\xspace}
\newcommand{\Fortran}{\textsf{Fortran}\xspace}
\newcommand{\Python}{\textsf{Python}\xspace}
\newcommand{\Ruby}{\textsf{Ruby}\xspace}

\newcommand{\AppArmor}{\texttt{AppArmor}\xspace}
\newcommand{\RAppArmor}{\pkg{RAppArmor}\xspace}
\newcommand{\Linux}{\texttt{Linux}\xspace}
\newcommand{\ULIMIT}{\texttt{ULIMIT}\xspace}

\author{Jeroen Ooms\\UCLA Department of Statistics} 

\title{The \RAppArmor Package: Enforcing Security Policies in \R Using Dynamic
Sandboxing on Linux}

\Plainauthor{Jeroen Ooms}  

\Plaintitle{The RAppArmor Package: Enforcing Security Policies in R using
Dynamic Sandboxing on Linux} 

\Shorttitle{The RAppArmor Package}

\Abstract{
  The increasing availability of cloud computing and scientific super computers
  brings great potential for making \R accessible through public or shared
  resources. This allows us to efficiently run code requiring lots of cycles and
  memory, or embed \R functionality into e.g.\ systems and web services. However
  some important security concerns need to be addressed before this can be put
  in production. The prime use case in the design of \R has always been a single
  statistician running \R on the local machine through the interactive console.
  Therefore the execution environment of \R is entirely unrestricted, which
  could result in malicious behavior or excessive use of hardware resources in a
  shared environment. Properly securing an \R process turns out to be a complex
  problem. We describe various approaches and illustrate potential issues using
  some of our personal experiences in hosting public web services. Finally we
  introduce the \RAppArmor package: a Linux based reference implementation for
  dynamic sandboxing in \R on the level of the operating system.
}

\Plainkeywords{R, security, linux, sandbox, apparmor}

\Address{
  Jeroen Ooms\\
  UCLA Department of Statistics\\
  University of California\\
  E-mail: \email{jeroen.ooms@stat.ucla.edu}\\
  URL: \url{http://jeroenooms.github.io}
}

\begin{document}

\section[Security in R: Introduction and motivation]{Security in \R:
Introduction and motivation}

The \R project for statistical computing and graphics \citep{R-project} is
currently one of the primary tool-kits for scientific computing. The software is
widely used for research and data analysis in both academia and industry, and is
the de-facto standard among statisticians for the development of new
computational methods. With support for all major operating systems, a powerful
standard library, over 3000 add-on packages and a large active community, it is
fair to say that the project has matured to a production-ready computation tool.
However, practices in statistical computation have changed since the initial
design of \R in 1993 \citep{ihaka1998r}. Internet access, public cloud computing
\citep{armbrust2010view}, live and open data and scientific super computers are
transforming the landscape of data analysis. This is only the beginning. Sharing
of data, code and results on social computing platforms will likely become an
integral part of the publication process \citep{asareport}. This could address
some of the hardware challenges, but also contribute towards reproducible
research and further socialize data analysis, i.e. facilitate learning,
collaboration and integration. These developments shift the role of
statistical software towards more general purpose computational back-ends,
powering systems and applications with embedded analytics and visualization.

However, one reason developers might still be reluctant to build on \R is
concerns regarding security and management of shared hardware resources.
Reliable software systems require components which behave predictably and
cannot be abused. Because \R was primarily designed with the local user in
mind, security restrictions and unpredictable behavior have not been considered
a major concern in the design of the software. Hence, these problems will need
to be addressed somehow before developers can feel comfortable making \R part
of their infrastructure, or convince administrators to expose their facilities
to the public for \R based services. It is our personal experience that the
complexity of managing security is easily underestimated when designing stacks
or systems that build on \R. Some of the problems are very domain specific to
scientific computing, and make embedding \R quite different from embedding
other software environments. Properly addressing these challenges can help
facilitate wider adoption of \R as a general purpose statistical engine.

\subsection{Security when using contributed code}

Building systems on \R has been the main motivation for this research. However,
security is a concern for \R in other contexts as well. As the community is
growing rapidly, relying on social courtesy in contributed code becomes more
dangerous. For example, on a daily basis, dozens of packages and package updates
submitted to the \emph{Comprehensive R Archive Network} (CRAN)
\citep{ripleycran}. These packages contain code written in \R, \C, \Fortran,
\Cpp, \Java, etc. It is unfeasible for the CRAN maintainers to do a thorough
audit of the full code that is submitted, every time. Some packages even contain
pre-compiled \Java code for which the source is not included. Furthermore, \R
packages are not signed with a private key as is the case for e.g.\ packages in
most \Linux distributions, which makes it hard to verify the identity of the
author. As CRAN packages are automatically build and installed on hundreds,
possibly thousands of machines around the world, they form an interesting
target for abuse. Hence there is a real dangler of packages containing
malicious code making their way unnoticed into the repositories. Risks are even
greater for packages distributed through channels without any form of code
review, for example by email or through the increasingly popular Github
repositories \citep{torvalds2010git,dabbish2012social}.

In summary, it is not overly paranoid of the \R user to be a bit cautious when
installing and running contributed code downloaded from the internet. However,
things don't have to be as critical as described above. Controlling security is
a good practice, even when there are no immediate reasons for concern. Some
users simply might want to prevent \R from accidentally erasing files or
interfering with other activities on the machine. Making an effort to ensure \R
is running safely with no unnecessary privileges can be reassuring to both user
and system administrator, and might one day prevent a lot of trouble.

\subsection[Sandboxing the R environment]{Sandboxing the \R environment}

This paper explores some of the potential problems, along with approaches and
methods of securing \R. Different aspects of security in the context of \R are
illustrated using personal experiences and examples of bad or malicious code. We
will explain how untrusted code can be executed inside a \emph{sandboxed}
process. Sandboxing in this context is a somewhat informal term for creating an
execution environment which limits capabilities of harmful and undesired
behavior. As it turns out, \R itself is not very suitable for implementing such
access control policies, and the only way to properly enforce security is by
leveraging features from the operating system. To exemplify this approach, an
implementation based on \AppArmor is provided which can be used on \Linux
distributions as the basis for a sandboxing toolkit. This package is used
throughout the paper to demonstrate one way of addressing issues.

However, we want to emphasize that we don't claim to have solved the problem.
This paper mostly serves an introduction to security for the \R community, and
hopefully creates some awareness that this is a real issue moving forward. The
\RAppArmor package is one approach and a good starting point for experimenting
with dynamic sandboxing in \R. However it mostly serves as a proof of concept of
the general idea of confining and controlling an \R process. The paper describes
examples of use cases, threats, policies and anecdotes to give the reader a
sense of what is involved with this topic. Without any doubt, there are concerns
beyond the ones mentioned in this paper, many of which might be specific to
certain applications or systems. We hope to invoke a discussion in the community about
potential security issues related to using \R in different scenarios, and
encourage those comfortable with other platforms or who use \R for different
purposes to join the discussion and share their concerns, experiences and
solutions.

\section[Use cases and concerns of sandboxing R]{Use cases and concerns of
sandboxing \R}

Let us start by taking a step back and put this research in perspective by
describing some concrete use cases where security in \R could be a concern.
Below three simple examples of situations in which running \R code in a sandbox
can be useful. The use cases are ordered by complexity and require increasingly
advanced sandboxing technology.

\subsubsection{Running untrusted code}

Suppose we found an \R package in our email or on the internet that looks
interesting, but we are not quite sure who the author is, and if the package
does not contain any malicious code. The package is too large for us to inspect
all of the code manually, and furthermore it contains a library in a foreign
language (e.g.\ \Cpp, \Fortran) for which we lack the knowledge and insight to
really understand its implications. Moreover, programming style (or lack
thereof) of the author can make it difficult to assess what exactly is going on
\citep{ioccc}. Nevertheless we would like to give the package a try, but without
exposing ourselves to the risk of potentially jeopardizing the machine.

One solution would be to run untrusted code on a separate or virtual machine.
We could either install some local virtualization software, or rent a VPS/cloud
server to run R remotely, for example on Amazon EC2. However this is somewhat
cumbersome and we will not have our regular workflow available: in order to put
the package to the test on our own data, we first need to copy our data,
scripts, files and package library, etc. Some local virtualization software can
be configured for read-only sharing of resources between host and guest machine,
but we would still need separate virtual machines for different tasks and
projects. In practice, managing multiple machines is a bit unpractical and not
something that we might want to do on a daily basis. It would be more convenient
if we could instead sandbox our regular \R environment for the duration of
installing and using the new package with a tailored security policy. If the
sandbox is flexible and unobtrusive enough not to interfere with our daily
workflow, we could even make a habit out of using it each time we use
contributed code (which to most users means every day).

\subsubsection{Shared resources}

A second use case could be a scenario where multiple users are sharing a single
machine. For example, a system administrator at a university is managing a large
computing server and would like to make it available to faculty and students.
This would allow them to run \R code that requires more computing power than
their local machine can handle. For example a researcher might want to do a
simulation study, and fit a complex model a million times on generated datasets
of varying properties. On her own machine this would take months to complete,
but the super computer can finish the job overnight. The administrator would
like to set up a web service for this and other researchers to run such R
scripts. However he is worried about users interfering with each others work, or
breaking anything on the machine. Furthermore he wants to make sure that system
resources are allocated in a fair way such that no single user can consume all
memory or cpu on the system.

\subsubsection{Embedded systems and services}

There have numerous efforts to facilitate integration of \R functionality into
3rd party systems, both for open source and proprietary purposes. Major
commercial vendors like Oracle, IBM and SAS have included \R interfaces in their
products. Examples of open source interfaces from popular general purpose
languages are \texttt{RInside} \citep{RInside}, which embeds \R into \Cpp
environments, and \texttt{JRI} which embeds \R in \Java software
\citep{JRI,urbanek2007rjava}. Similarly, \texttt{rpy2}
\citep{moreira2006rpy,gautier2008rpy2} provides a \Python interface to \R, and
\texttt{RinRuby} is a \Ruby library that integrates the \R interpreter in Ruby
\citep{dahl2008rinruby}. Littler provides hash-bang (i.e.\ script starting with
\texttt{\#!/some/path}) capability for \R \citep{littler}. The Apache2 module
rApache (\texttt{mod\_R}) \citep{rapache} makes it possible to run \R scripts
from within the Apache2 web server. \cite{heiberger2009r} provide a series of
tools to call \R from DCOM clients on Windows environments, mostly to support
calling \R from Microsoft Excel. Finally, \texttt{RServe} is TCP/IP server which
provides low level access to an \R session over a socket \citep{Rserve}.

The third use case originates from these developments: it can be summarized as
confining and managing \R processes inside of embedded systems and services.
This use case is largely derived from our personal needs: we are using \R inside
various systems and web services to provide on-demand calculating and plotting
over the internet. These services need to respond quickly and with minimal
overhead to incoming requests, and should scale to serve many jobs per second.
Furthermore the systems need to be stable, requiring that jobs should always
return within a given timeframe. Depending on user and the type of job,
different security restrictions might be appropriate. Some services specifically
allow for execution of arbitrary \R code. Also we need to dynamically enforce
limits on the use of memory, processors and disk space on a per process basis.
These requirements demand a more flexible and finer degree of control over the
process privileges and restrictions than the first two use cases. It encouraged
us to explore more advanced methods than the conventional tools and has been
the most central motivation of this research.

\subsection{System privileges and hardware resources}

The use cases described above outline motivations and requirements for an \R
sandbox. Two inter-related problems can be distinguished. The first one is
preventing system abuse, i.e.\ use of the machine for malicious or undesired
activity, or completely compromising the machine. The second problem is managing
hardware resources, i.e.\ preventing excessive use by constraining the amount of
memory, cpu, etc that a single user or process is allowed to consume.

\subsubsection{System abuse}

The \R console gives the user direct access to the operating system and does not
implement any privilege restrictions or access control policies to prevent
malicious use. In fact, some of the basic functionality in \R assumes quite
profound access to the system, e.g.\ read access to system files, or the
privilege of running system shell commands. However, running untrusted \R code
without any restrictions can get us in serious trouble. For example, the code
could call the \texttt{system()} function from where any shell commands can be
executed. But also innocent looking functions like \texttt{read.table} can be
used to extract sensitive information from the system, e.g.\
\texttt{read.table("/etc/passwd")} lists all users on the system or
\texttt{readLines("/var/log/syslog")} exposes system log information.

Even an \R process running as a non-privileged user can do a lot of harm.
Potential perils include code containing or downloading a virus or security
exploit, or searching the system for sensitive personal information. Appendix
\ref{creditcard} demonstrates a hypothetical example of a simple function that
scans the home directory for documents containing credit card numbers. Another
increasing global problem are viruses which make the machine part of a so called
``botnet''. Botnets are large networks of compromised machines (``bots'') which
are remotely controlled to used for illegal activities
\citep{abu2006multifaceted}. Once infected, the botnet virus connects to a
centralized server and waits for instructions from the owner of the botnet.
Botnets are mostly used to send spam or to participate in DDOS attacks:
centrally coordinated operations in which a large number of machines on the
internet is used to flood a target with network traffic with the goal of taking
it down by overloading it \citep{mirkovic2004taxonomy}. Botnet software is often
invisible to the user of an infected machine and can run with very little
privileges: simple network access is sufficient to do most of its work.

When using \R on the local machine and only running our own code, or from
trusted sources, these scenarios might sound a bit far fetched. However, when
running code downloaded from the internet or exposing systems to the public,
this is a real concern. Internet security is a global problem, and there are a
large number of individuals, organizations and even governments actively
employing increasingly advanced and creative ways of gaining access to protected
infrastructures. Especially servers running on beefy hardware or fast
connections are attractive targets for individuals that could use these
resources for other purposes. But also servers and users inside large companies,
universities or government institutions are frequently targeted with the goal of
gathering confidential information. This last aspect seems especially relevant,
as \R is used frequently in these types of organizations.

\subsubsection{Resource restrictions}

The other category of problems is not necessarily related to deliberate abuse,
and might even arise completely unintentionally. It involves proper management,
allocation and restricting of hardware.

It is fair to say that \R can be quite greedy with system resources. It is easy
to run a command which will consume all of the available memory and/or CPU, and
does not finish executing until manually terminated. When running \R on the
local machine through the interactive console, the user will quickly recognize a
function call that is not returning timely or is making the machine
unresponsive. When this happens, we can easily interrupt the process prematurely
by sending a \texttt{SIGINT}, i.e.\ pressing \texttt{CTRL+C} in \Linux or
\texttt{ESC} in Windows. If this doesn't work we can open the task manager and
tell the operating system to kill the process, and if worst comes to worst we
can decide to reboot our machine.

However, when \R is embedded in a system, the situation is more complicated and
we need to cover these scenarios in advance. If an out-of-control \R job is not
properly detected and terminated, the process might very well run indefinitely
and take down our service, or even the entire machine. This has actually been a
major problem that we personally experienced in an early implementation of a
public web service for mixed modelling \citep{yeroonlme4} which uses the
\texttt{lme4} package \citep{lme4}. What happened was that users could
accidentally specify a variable with many levels as the \emph{grouping factor}.
This causes the design matrix to blow up even on a relatively small dataset, and
decompositions will take forever to complete. To make things worse,
\texttt{lme4} uses a lot of \texttt{C} code which does not respond to time
limits set by R's \texttt{setTimeLimit} function. Appendix \ref{cputime}
contains a code snippet that simulates this scenario. When this would happen,
the only way to get things up and running again was to manually login to the
server and reset the application.

Unfortunately this example is not an exception. The behavior of \R can be
unpredictable, which is an aspect easily overlooked by (non-statistician)
developers. When a system calls out to e.g.\ an \texttt{SQL} or \texttt{PHP}
script, the procedure usually runs without any problems and the processing time
is proportional to the size of the data, i.e.\ the number of records returned by
\texttt{SQL}. However, in an \R script many things can go wrong, even though the
script itself is perfectly fine.  Algorithms might not converge, data might be
rank-deficient, or missing values throw a spanner in the works.
Statistical computing is full of such intrinsic edge-cases and unexpected
caveats.
Using only tested code or predefined services does not entirely guarantee smooth
and timely completion of \R jobs, especially if the data is dynamic. When
embedding \R in systems or shared facilities, it is important that we
acknowledge this facet and have systems in place to manage jobs and mitigate
problems without manual intervention.

\section[Various approaches of confining R]{Various approaches of confining
\R}

The current section introduces several approaches of securing and sandboxing \R,
with their advantages and limitations. They are reviewed in the context of our
use cases, and evaluated on how they address the problems of system abuse and
restricting resources. The approaches are increasingly \emph{low-level}: they
represent security on the level of the application, R software itself and
operating system respectively. As will become clear, we are leaning towards the
opinion that \R is not very well suited to address security issues, and the only
way to do proper sandboxing is on the level of the operating system. This will
lead up to the \RAppArmor package introduced in section \ref{rapparmor}.

\subsection{Application level security: predefined services}

The most common approach to prevent malicious use is simply to only allow a
limited set of predefined services, that have been deployed by a trusted
developer and cannot be abused. This is generally the case for websites
containing dynamic content though e.g.\ \proglang{CGI} or \proglang{PHP}
scripts.
Running arbitrary code is explicitly prevented and any possibility to do so
anyway is considered a security hole. For example, we might expose the following
function as a web service:

\begin{CodeChunk}
\begin{CodeInput} 
liveplot <- function (ticker) {
  url <- paste("http://ichart.finance.yahoo.com/table.csv?s=",
    ticker, "&a=07&b=19&c=2004&d=07&e=13&f=2020&g=d&ignore=.csv",
    sep = "")
  mydata <- read.csv(url)
  mydata$Date <- as.Date(mydata$Date)
  myplot <- ggplot2::qplot(Date, Close, data = mydata, geom = c("line",
    "smooth"), main = ticker)
  print(myplot)
}
\end{CodeInput}
\end{CodeChunk}

The function above downloads live data from the public API at Yahoo Finance and
creates an on-demand plot of the historical prices using \texttt{ggplot2}
\citep{ggplot2}. It has only one parameter: \texttt{ticker},
a character string identifying a stock symbol. This function can be exposed as a
predefined web service, where the client only supplies the \texttt{ticker}
argument. Hence the system does not need to run any potentially harmful
user-supplied \R code. The client sets the symbol to e.g.
\texttt{"GOOG"} and the resulting plot can be returned in the form of a
PNG image or PDF document. This function is actually the basis of the
``stockplot'' web application \citep{stockplot}; an interactive graphical web
application for financial analysis which still runs today.

Limiting users and clients to a set of predefined parameterized services is the
standard solution and reasonably safe in combination with basic security
methods. For example, \pkg{Rserve} can be configured to run with a custom
\texttt{uid}, \texttt{umask}, \texttt{chroot}. However in the context of \R,
predefined services severely limit the application and security is actually not
fully guaranteed. We can expose some canned calculations or generate a plot as
done in the example, but beyond that things quickly becomes overly restrictive.
For example in case of an application that allows the user to fit a statistical
model, the user might need to be able to include transformations of variables
like \texttt{I(cos(x\^\ 2))} or \texttt{cs(x, 3)}. Not allowing a user to call
any custom functions makes this hard to implement.

What distinguishes \R from other statistical software is that the user has a
great deal of control and can do programming, load custom libraries, etc. A
predefined service takes this freedom away from the user, and at the same time
puts a lot of work in the hands of the developer and administrator. Only they
can expose new services and they have to make sure that all services that are
exposed cannot be abused in some way or another. Therefore this approach is
expensive, and not very social in terms of users contributing code. In
practice, anyone that wants to publish an \R service will have to purchase and
manage a personal server or know someone that is willing to do so. Also it is
still important to set hardware limitations, even when exposing relatively
simple, restricted services. We already mentioned the example of the
\texttt{lme4} web application, where a single user could accidentally take down
the entire system by specifying an overly complex model. But actually even some
of the most basic functionality in \R can cause trouble with problematic data.
Hence, even restricted predefined \R services are not guaranteed to
consistently return smooth and timely. These aspects of statistical computing
make common practices in software design not directly generalize to \R
services, and are easily under appreciated by developers and engineers with a
limited background in data analysis.

\subsubsection{Code injection}

Finally, there is still the risk of \emph{code injection}. Because \R is a very
dynamic language, evaluations sometimes happen at unexpected places. One example
is during the parsing of \texttt{formulas}. For example, we might want to
publish a service that calls the \texttt{lm()} function in \R on a fixed
dataset. Hence the only parameter supplied by the user is a \emph{formula} in
the form of a character vector. Assume in the code snippet below that the
 \texttt{userformula} parameter is a string that has been set through some
graphical interface.

\begin{CodeChunk}
\begin{CodeInput}
coef(lm(userformula, data=cars))
\end{CodeInput}
\end{CodeChunk}

For example the user might supply a string
\texttt{"speed{\raise.17ex\hbox{$\scriptstyle\sim$}}dist"} and the service
will return the coefficients. On first sight, this might seem like a safe
service. However, formulas actually allow for the inclusion of calls to other
functions. So even though \texttt{userformula} is a character vector, it can be
used to inject a function call:

\begin{CodeChunk}
\begin{CodeInput}
userformula <- "speed ~ dist + system('whoami')"
lm(userformula, data=cars)
\end{CodeInput}
\end{CodeChunk}

In the example above, \texttt{lm} will automatically convert
\texttt{userformula} from type character to a \texttt{formula}, and subsequently
execute the \texttt{system('whoami')} command. Hence even when a client supplies
only simple primitive data, unexpected opportunities for code injection can
still arise. Therefore it is important when using this approach, to sanitize the
input before executing the service. One way is by setting up the service such
that only alphanumeric values are valid parameters, and use a regular
expression to remove any other characters, before actually executing the script
or service:
\begin{CodeChunk}
\begin{CodeInput}
myarg <- gsub("[^a-zA-Z0-9]", "", myarg)
\end{CodeInput}
\end{CodeChunk}

\subsection{Sanitizing code by blacklisting}

A less restrictive approach is to allow users to push custom \R code, but
inspect the code before evaluating it. This approach has been adopted by some
web sites that allow users to run \R code, like \cite{banfield1999rweb} and
\cite{cloudstat}. However, given the dynamic nature of the \R language,
malicious calls are actually very difficult to detect and such security is easy
to circumvent. For example, we might want to prevent users from calling the
\texttt{system} function. One way is to define some smart regular expressions
that look for the word ``system'' in a block of code. This way it would be
possible to detect a potentially malicious call like this:

\begin{CodeChunk}
\begin{CodeInput}
system("whoami")
\end{CodeInput}
\end{CodeChunk}

However, it is much more difficult to detect the equivalent call in the following
block:

\begin{CodeChunk}
\begin{CodeInput}
foo <- get(paste("sy", "em", sep="st"))
bar <- paste("who", "i", sep="am")
foo(bar)
\end{CodeInput}
\end{CodeChunk}

And indeed, it turns out that the services that use this approach are fairly
easy to trick. Because \R is a dynamic scripting language, the exact function
calls might not reveal themselves until runtime, when it is often too late. We
are actually quite convinced that it is nearly impossible to really sanitize an
\R script just by inspecting the source code.

An alternative method to prevent malicious code is by defining an extensive
blacklist of functions that a user is not allowed to call, and disable these at
runtime. The \pkg{sandboxR} package \citep{sandboxR}  does this to block access
to all \R functions providing access to the file system. It evaluates the
user-supplied code in an environment in which all blacklisted functions are
masked from the calling namespace. This is fairly effective and provides a
barrier against smaller possible attacks or casual errors. However, the method
relies on exactly knowing and specifying which functions are \emph{safe} and
which are not. The package author has done this for the thousands of \R
functions in the base package and we assume he has done a good job. But it is
quite hard to maintain and cumbersome to generalize to other \R packages (by
default the method does not allow loading other packages). Everything falls if
one function has been overlooked or changes between versions, which does make
the method vulnerable. Furthermore, in \R even the most primitive functions can
be exploited to tamper with scoping and namespaces, so it is unwise to rely
solely on this for security.

Moreover, even when sanitizing of the code is successful, this method does not
limit the use of hardware resources in any way. Hence, additional methods are
still required to prevent excessive use of resources in a public environment.
Packages like \pkg{sandboxR} should probably only be used to supplement system
level security as implemented in the \RAppArmor package. They can be useful to
detect problematic calls earlier on and present informative errors naming a
specific forbidden function rather than just "permission denied". But
blacklisting solutions are not waterproof and should not be considered a full
security solution.

\subsection{Sandboxing on the level of the operating system}

One can argue that managing resources and privileges is something that is
outside the domain of the \R software, and is better left to the operating
system. The \R software has been designed for statistical computing and related
functionality; the operating system deals with hardware and security related
matters. Hence, in order to really sandbox \R properly without imposing
unnecessary limitations on its functionality, we need to sandbox the
\emph{process} on the level of the operating system. When restrictions are
enforced by the operating system instead of \R itself, we do not have to worry
about all of the pitfalls and implementation details of \R. The user can
interact freely with \R, but won't be able to do anything for which the system
does not grant permission.

Some operating systems offer more advanced capabilities for setting process
restrictions than others. The most advanced functionality is found in
\texttt{UNIX} like systems, of which the most popular ones are either
\texttt{BSD} based (\texttt{FreeBSD, OSX,} etc) or \Linux based (\texttt{Debian,
Ubuntu, Fedora, Suse,} etc). Most \texttt{UNIX} like systems implement some sort
of \ULIMIT functionality to facilitate restricting availability of hardware
resources on a per-process basis. Furthermore, both \texttt{BSD} and \Linux
provide various \emph{Mandatory Access Control} (MAC) systems. On \Linux, these
are implemented as Kernel modules. The most popular ones are \AppArmor
\citep{apparmor}, \texttt{SELinux} \citep{selinux} and \texttt{Tomoyo Linux}
\citep{tomoyo}. MAC gives a much finer degree of control than standard
user-based privileges, by applying advanced security policies on a per-process
basis. Using a combination of MAC and \ULIMIT tools we can do a pretty decent
job in sandboxing a single \R process to a point where it can do little harm to
the system. Hence we can run arbitrary \R code without losing any sleep over
potentially jeopardizing the machine. Unfortunately, this approach comes at the
cost of portability of the software. As different operating systems implement
very different methods for managing processes and privileges, the solutions
will be to a large extend OS-specific. In our implementation we have tried to
hide these system calls by exposing \R functions to interact with the kernel.
Going forward, eventually these functions could behave somewhat OS specific,
abstracting away technicalities and providing similar functionality on
different systems. But for now we limit ourselves to systems based on the
\Linux kernel.

\section[The RAppArmor package]{The \RAppArmor package}
\label{rapparmor}

The current section describes some security concepts and how an \R process can
be sandboxed using a combination of \ULIMIT and \texttt{MAC} tools. The methods
are illustrated using the \RAppArmor package: an implementation based on \Linux
and \AppArmor. \AppArmor (``Application Armor'') is a security module for the
\Linux kernel. It allows for associating programs and processes with
\emph{security profiles} that restrict the capabilities and permissions of that
process. There are two ways of using \AppArmor. One is to associate a single,
static security profile with every \R process. This can be done only by the
system administrator and does not require our \R package (see also section
section \ref{usr.bin.r}). However, this is usually overly restrictive. We want
more flexibility to set different policies, priorities and restrictions for
different users or tasks.

The \RAppArmor package exposes \R functions that interface directly to \Linux
system calls related to setting privileges and restrictions of a running
process. Besides applying security profiles, \RAppArmor also interfaces to the
\texttt{prlimit} call in \Linux, which sets \texttt{RLIMIT} (resource limit)
values on a process (\texttt{RLIMIT} are the \Linux implementation of \ULIMIT).
\Linux defines a number of \texttt{RLIMIT}'s, which restrict resources like
memory use, number of processes, and stack size. More details on
\texttt{RLIMIT} follow in section \ref{RLIMITS}. Using \RAppArmor, the
sandboxing functionality is accessible directly from within the \R session,
without the need for external tools or programs. Once \RAppArmor is installed,
any user can apply security profiles and restrictions to the running process;
no special permissions are required. Furthermore, it allowed us to create the
\texttt{eval.secure} function: an abstraction which mimics \texttt{eval}, but
has additional parameters to evaluate a single call under a given uid,
priority, security policy and resource restrictions.

The \RAppArmor package brings the low level system security methods all the way
up to level of the \R language. Using \texttt{eval.secure}, different parts of
our code can run with different security restrictions with minimal overhead,
something we call \emph{dynamic sandboxing}. This is incredibly powerful in the
context of embedded services, and opens the door applications which explicitly
allow for arbitrary code execution; something that previously always had to be
avoided for security reasons. This enables a new approach to socialize
statistical computing and lies at the core of the \texttt{OpenCPU} framework
\citep{opencpu}, which exposes a public HTTP API to run and share \R code on a
central server.

\subsection[AppArmor profiles]{\AppArmor profiles}

Security policies are defined in \emph{profiles} which form the core of the
\AppArmor software. A profile consists of a set of rules specified using
\AppArmor syntax in an ascii file. The \Linux kernel translates these rules to
a security policy that it will enforce on the appropriate process. A brief
introduction to the \AppArmor syntax is given in section \ref{syntax}. The
appendix of this paper contains some example profiles that ship with the
\RAppArmor package to get the user started. When the package is installed
through the Debian/Ubuntu package manager (e.g.\ using \texttt{apt-get}) the
profiles are automatically copied to \texttt{/etc/apparmor.d/rapparmor.d/}.
Because profiles define file access permissions based the location of files and
directories on the file system, they are to some extent specific to a certain
\Linux distribution, as different distributions have somewhat varying
conventions on where files are located. The example profiles included with
\RAppArmor are based on the file layout of the \pkg{r-base} package (and its
dependencies) by \cite{batesusing} for Debian/Ubuntu, currently maintained by
Dirk Eddelbuettel.

The \texttt{RAppArmor} package and the included profiles work ``out of the box''
on Ubuntu 12.04 (Precise) and up, Debian 7.0 (Wheezy) and up. The package can
also be used on OpenSuse 12.1 and up, however Suse systems organize the file
system in a slightly different way than Ubuntu and Debian, so the profiles need
to be modified accordingly. The \RAppArmor website contains some specific
instructions regarding various distributions.

Again, we want to emphasize that the package should mostly be seen as a
\emph{reference implementation} to demonstrate how to create a working sandbox
in \R. The \RAppArmor package provides the tools to set security restrictions
and example profiles to get the user started. However, depending on system and
application, different policies might be appropriate. It is still up to the
administrator to determine which privileges and restrictions are appropriate
for a certain system or purpose. The example profiles are merely a starting
point and need fine-tuning for specific applications.

\subsection{Automatic installation}

The \RAppArmor package consists of \R software and a number of example security
profiles. On Ubuntu it is easiest installed using binary builds provided
through launchpad:

\begin{CodeChunk}
\begin{CodeInput}
sudo add-apt-repository ppa:opencpu/rapparmor
sudo apt-get update
sudo apt-get install r-cran-rapparmor
\end{CodeInput}
\end{CodeChunk}

Binaries in this repository are build for the version of \R that ships with the
operating system. To get builds for \R version 3.0 and up, use repository
\texttt{ppa:opencpu/rapparmor-dev} instead. The \texttt{r-cran-rapparmor}
package can also be build from source using something along the lines of the
following:

\begin{CodeChunk}
\begin{CodeInput}
wget http://cran.r-project.org/src/contrib/RAppArmor_0.8.3.tar.gz
tar xzvf RAppArmor_0.8.3.tar.gz
cd RAppArmor/
debuild -uc -us
cd ..
sudo dpkg -i r-cran-rapparmor_0.8.3-precise1_amd64.deb
\end{CodeInput}
\end{CodeChunk}

The \texttt{r-cran-rapparmor} package will automatically install required
dependencies and security profiles. The security profiles are installed in
\texttt{/etc/appamor.d/rapparmor.d/}.

\subsection{Manual installation}

On distributions for which no system installation package is available, manual
installation is required. Start with installing required dependencies:

\begin{CodeChunk}
\begin{CodeInput}
sudo apt-get install r-base-dev libapparmor-dev apparmor apparmor-utils
\end{CodeInput}
\end{CodeChunk}

Note that \R version 2.14 or higher is required. Also the system needs to have
an \AppArmor enabled \Linux kernel. After dependencies have been installed,
install \RAppArmor from CRAN:

\begin{CodeChunk}
\begin{CodeInput}
wget http://cran.r-project.org/src/contrib/RAppArmor_0.8.3.tar.gz
sudo R CMD INSTALL RAppArmor_0.8.3.tar.gz
\end{CodeInput}
\end{CodeChunk}

This will compile the \proglang{C} code and install the \R package. After the
package has been installed successfully, the security profiles need to be
copied to the \texttt{apparmor.d} directory:

\begin{CodeChunk}
\begin{CodeInput}
cd /usr/local/lib/R/site-library/RAppArmor/
sudo cp -Rf profiles/debian/* /etc/apparmor.d/
\end{CodeInput}
\end{CodeChunk}

Finally, the \AppArmor service needs to be restarted to load the new profiles.
Also we do not want to enforce the global \R profile at this point:

\begin{CodeChunk}
\begin{CodeInput}
sudo service apparmor restart
sudo aa-disable usr.bin.r
\end{CodeInput}
\end{CodeChunk}

This should complete the installation. To verify if everything is working, start
\R and run the following code:

\begin{CodeChunk}
\begin{CodeInput}
library("RAppArmor")
aa_change_profile("r-base")
\end{CodeInput}
\end{CodeChunk}

If the code runs without any errors, the package has successfully been
installed. 

\subsection[Linux security methods]{\Linux security methods}

The \RAppArmor package interfaces to a number of \Linux system calls that are
useful in the context of security and sandboxing. The advantage of calling
these directly from \R is that we can dynamically set the parameters from
within the \R process, as opposed to fixing them for all \R sessions. Hence it
is actually possible to execute some parts of an application in a different
security context other parts.

The package implements many functions that wrap around \Linux \proglang{C}
interfaces. However it is not required to study all of these functions. To the
end user, everything in the package comes together in the powerful and
convenient \texttt{eval.secure()} function. This function mimics
\texttt{eval()}, but it has additional parameters that define restrictions
which will be enforced to a specific evaluation. An example:

\begin{CodeChunk}
\begin{CodeInput}
myresult <- eval.secure(myfun(), RLIMIT_AS = 10*1024*1024, profile="r-base")
\end{CodeInput}
\end{CodeChunk}

This will call \texttt{myfun()} with a memory limit of 10MB and the ``r-base''
security profile (which is introduced in section \ref{r-base-intro}). The
\texttt{eval.secure} function works by creating a \emph{fork} of the current
process, and then sets hard limits, \texttt{uid} and \AppArmor profile on the
forked process, before evaluating the call. After the function returns, or when
the timeout is reached, the forked process is killed and cleaned up. This way,
all of the one-way security restrictions can be applied, and evaluations that
happen inside \texttt{eval.secure} won't have any side effects on the main
process.

\subsection{Setting user and group ID}

One of the most basic security methods is running a process as a specific user.
Especially within a system where the main process has superuser privileges
(which could be the case in for example a webserver), switching to a user with
limited privileges before evaluating any code is a wise thing to do. We could
even consider a design where every user of the application has a dedicated
user account on the \Linux machine. The \RAppArmor package implements the
functions \texttt{getuid, setuid, getgid, setgid}, which call out to the
respective \Linux system calls. Users and groups can either be specified
by their name, or as integer values as defined in the \texttt{/etc/passwd} file.

\begin{CodeChunk}
\begin{CodeInput}
R> library("RAppArmor")
AppArmor LSM is enabled.
Current profile: none (unconfined).
R> system('whoami')
root
R> getuid()
[1] 0
R> getgid()
[1] 0
R> setgid(1000)
R> setuid(1000)
R> getgid()
[1] 1000
R> getuid()
[1] 1000
R> system('whoami')
jeroen
\end{CodeInput}
\end{CodeChunk}

The user/group ID can also be set inside the \texttt{eval.secure} function. In
this case it will not affect the main process; the UID is only set for the time
of the secure evaluation.

\begin{CodeChunk}
\begin{CodeInput}
R> eval(system('whoami', intern=TRUE))
[1] "root"
R> eval.secure(system('whoami', intern=TRUE), uid=1000)
[1] "jeroen"
R> eval(system('whoami', intern=TRUE))
[1] "root"
\end{CodeInput}
\end{CodeChunk}

Note that in order for \texttt{setgid} and \texttt{setuid} to work, the user
must have the appropriate capabilities in \Linux, which are usually
restricted to users with superuser privileges. The \texttt{getuid} and
\texttt{getgid} functions can be called by anyone.

\subsection{Setting Task Priority}
\label{priority}

The \RAppArmor package implements interfaces for setting the scheduling priority
of a process, also called its \texttt{nice} value or \emph{niceness}. \Linux
systems use a priority system with 40 priorities, ranging from -20 (highest
priority) to 19 (lowest priority). By default most processes run with nice
value 0. Users without superuser privileges can only increase this value, i.e.
lower the priority of the process. In \RAppArmor the \texttt{getpriority} and
\texttt{setpriority} functions change the priority of the current session:

\begin{CodeChunk}
\begin{CodeInput}
R> getpriority()
[1] 0
R> setpriority(10)
[1] 10
R> system("nice", intern=TRUE)
[1] "10"
R> setpriority(5)
Error in setpriority(5) : 
  Failed to set priority. The caller attempted to lower a process priority, 
  but did not have the required privilege.
\end{CodeInput}
\end{CodeChunk}

Again, the \texttt{eval.secure} function is used to run a function or code block
with a certain priority without affecting the priority of the main \R session:

\begin{CodeChunk}
\begin{CodeInput}
R> getpriority()
[1] 0
R> eval.secure(system('nice', intern=TRUE), priority=10)
[1] "10"
R> getpriority()
[1] 0
\end{CodeInput}
\end{CodeChunk}

\subsection{Linux Resource Limits (RLIMIT)}
\label{RLIMITS}

Linux defines a number of \texttt{RLIMIT} values that can be used to set
resource limits on a process \citep{linuxrlimit}. The \RAppArmor package
has functions to get/set to the following RLIMITs:

\begin{itemize}
  \item \texttt{RLIMIT\_AS} -- The maximum size of the process's virtual memory
  (address space).
  \item \texttt{RLIMIT\_CORE} -- Maximum size of core file.
  \item \texttt{RLIMIT\_CPU} -- CPU time limit.
  \item \texttt{RLIMIT\_DATA} --  The maximum size of the process's data
  segment.
  \item \texttt{RLIMIT\_FSIZE} --  The maximum size of files that the process
  may create.
  \item \texttt{RLIMIT\_MEMLOCK} -- Number of memory that may be locked into
  RAM.
  \item \texttt{RLIMIT\_MSGQUEUE} -- Max number of bytes that can be allocated
  for POSIX message queues
  \item \texttt{RLIMIT\_NICE} --  Specifies a ceiling to which the process's
  nice value (priority).
  \item \texttt{RLIMIT\_NOFILE} -- Limit maximum file descriptor number that
  can be opened.
  \item \texttt{RLIMIT\_NPROC} -- Maximum number of processes (or, more
  precisely on Linux, threads) that can be created by the user of the calling
  process.
  \item \texttt{RLIMIT\_RTPRIO} -- Ceiling on the real-time priority that may be
  set for this process.
  \item \texttt{RLIMIT\_RTTIME} -- Limit on the amount of CPU time that a
  process scheduled under a real-time scheduling policy may consume without
  making a blocking system call.
  \item \texttt{RLIMIT\_SIGPENDING} -- Limit on the number of signals that may
  be queued by the user of the calling process.
  \item \texttt{RLIMIT\_STACK} -- The maximum size of the process stack.
\end{itemize}

For all of the above \texttt{RLIMITs}, the \RAppArmor package implements a
function which name is equivalent to the non-capitalized name of the
\texttt{RLIMIT}. For example to get/set \texttt{RLIMIT\_AS}, the user calls
\texttt{rlimit\_as()}. Every \texttt{rlimit\_} function has exactly 3 parameters:
\texttt{hardlim}, \texttt{softlim}, and \texttt{pid}. Each argument is
specified as an integer value. The \texttt{pid} arguments points to the target
process. When this argument is omitted, the calling process is targeted. When
the \texttt{softlim} is omitted, it is set equal to the \texttt{hardlim}.
When the function is called without any arguments, it returns the current limits.

The soft limit is the value that the kernel enforces for the corresponding
resource. The hard limit acts as a ceiling for the soft limit: an unprivileged
process may only set its soft limit to a value in the range from 0 up to the
hard limit, and (irreversibly) lower its hard limit. A  privileged process
(under  \Linux:  one  with  the \texttt{CAP\_SYS\_RESOURCE} capability) may make
arbitrary changes to either limit value.  \citep{linuxrlimit}

\begin{CodeChunk}
\begin{CodeInput}
R> rlimit_as()
$hardlim
[1] 1.845e+19

$softlim
[1] 1.845e+19

R> A <- rnorm(1e7)
R> rm(A)
R> gc()
         used (Mb) gc trigger (Mb) max used (Mb)
Ncells 230219  6.2     467875 12.5   350000  9.4
Vcells 211018  1.7    8773042 67.0 10211897 78.0
R> rlimit_as(10*1024*1024)
$hardlim
[1] 10485760

$softlim
[1] 10485760

R> A <- rnorm(1e7)
Error: cannot allocate vector of size 76.3 Mb
\end{CodeInput}
\end{CodeChunk}

Note that a process owned by a user without superuser privileges can only modify
\texttt{RLIMIT} to more restrictive values. However, using \texttt{eval.secure},
a more restrictive \texttt{RLIMIT} can be applied to a single evaluation without
any side effects on the main process:

\begin{CodeChunk}
\begin{CodeInput}
R> library("RAppArmor")
R> A <- eval.secure(rnorm(1e7), RLIMIT_AS = 10*1024*1024)
Error: cannot allocate vector of size 76.3 Mb
R> A <- rnorm(1e7)
\end{CodeInput}
\end{CodeChunk}

The exact meaning of the different limits can be found in the \texttt{RAppArmor}
package documentation (e.g.\ \texttt{?rlimit\_as}) or in the documentation of
the distribution, e.g.\ \cite{ubunturlimit}.

\subsection[Activating AppArmor profiles]{Activating \AppArmor profiles}

The \RAppArmor package implements three calls to the \Linux kernel related
to applying \AppArmor profiles: \texttt{aa\_change\_profile},
\texttt{aa\_change\_hat} and \texttt{aa\_revert\_hat}. Both the
\texttt{aa\_change\_profile} and \texttt{aa\_change\_hat} functions take a
parameter named \texttt{profile}: a character string identifying the name of
the profile. This profile has to be preloaded by the kernel, before it can be
applied to a process. The easiest way to load profiles is to copy them to the
directory \texttt{/etc/apparmor.d/} and then run \texttt{sudo service apparmor
restart}.

The main difference between a \emph{profile} and a \emph{hat} is that switching
profiles is an irreversible action. Once the profile has been associated with
the current process, the process cannot call \texttt{aa\_change\_profile} again
to escape from the profile (that would defeat the purpose). The only exception
to this rule are profiles that contain an explicit \texttt{change\_profile}
directive. The \texttt{aa\_change\_hat} function on the other hand is designed to
associate a process with a security profile in a way that does allow it to escape
out of the security profile. In order to realize this, the
\texttt{aa\_change\_hat} takes a second argument called \texttt{magic\_token},
which defines a secret key that can be used to \emph{revert} the hat. When
\texttt{aa\_revert\_hat} is called with the same \texttt{magic\_token} that
was used in \texttt{aa\_change\_hat}, the security restrictions are
relieved.

Using \texttt{aa\_change\_hat} to switch in and out of profiles is an easy way
to get started with \RAppArmor and test some security policies. However it
should be emphasized that using \emph{hats} instead of \emph{profiles} is also a
security risk and should be avoided in production settings. It is important to
realize that if the code running in the sandbox can find a way of discovering
the value of the \texttt{magic\_token} (e.g.\ from memory, command history or log
files), it will be able to escape from the sandbox. Hence
\texttt{aa\_change\_hat} should only be used to prevent general purpose
malicious activity, e.g.\ when testing a new \R package. When hosting services
or otherwise exposing an environment that might be specifically targeted,
hackers could write code that attempts to find the magic token and revert the
hat. Therefore it is recommended to only use \texttt{aa\_change\_profile} or
\texttt{eval.secure} in production settings. When a profile is applied to a
process using \texttt{aa\_change\_profile} or \texttt{eval.secure}, the kernel
will keep enforcing the security policy on the respective process and all of its
children until they die, no matter what.

The \RAppArmor package ships with a profile called \emph{testprofile} which
contains a hat called \emph{testhat}. We use this profile to demonstrate the
functionality. The profiles have been defined such that \emph{testprofile}
allows access to \texttt{/etc/group} but denies access to \texttt{/etc/passwd}.
The \emph{testhat} denies access to both \texttt{/etc/passwd} and
\texttt{/etc/group}.

\begin{CodeChunk}
\begin{CodeInput}
R> library("RAppArmor");
R> result <- read.table("/etc/passwd")

R> aa_change_profile("testprofile")
Switching profiles...
R> passwd <- read.table("/etc/passwd")
Error in file(file, "rt") : cannot open the connection
In addition: Warning message:
In file(file, "rt") : cannot open file '/etc/passwd': Permission denied
R> group <- read.table("/etc/group")

R> mytoken <- 13337;
R> aa_change_hat("testhat", mytoken)
Setting Apparmor Hat...

R> passwd <- read.table("/etc/passwd")
Error in file(file, "rt") : cannot open the connection
In addition: Warning message:
In file(file, "rt") : cannot open file '/etc/passwd': Permission denied
R> group <- read.table("/etc/group")
Error in file(file, "rt") : cannot open the connection
In addition: Warning message:
In file(file, "rt") : cannot open file '/etc/group': Permission denied

R> aa_revert_hat(mytoken);
Reverting AppArmor Hat...

R> passwd <- read.table("/etc/passwd")
Error in file(file, "rt") : cannot open the connection
In addition: Warning message:
In file(file, "rt") : cannot open file '/etc/passwd': Permission denied
R> group <- read.table("/etc/group")
\end{CodeInput}
\end{CodeChunk}

Just like for \texttt{setuid} and \texttt{rlimit} functions,
\texttt{eval.secure} can be used to enforce an \AppArmor security profile on a
single call, witout any side effects. The \texttt{eval.secure} function uses
\texttt{aa\_change\_profile} and is therefore most secure.

\begin{CodeChunk}
\begin{CodeInput}
R> out <- eval(read.table("/etc/passwd"))
R> nrow(out)
[1] 68
R> out <- eval.secure(read.table("/etc/passwd"), profile="testprofile")
Error in file(file, "rt") : cannot open the connection
\end{CodeInput}
\end{CodeChunk}

\subsection[AppArmor without RAppArmor]{\AppArmor without \RAppArmor}
\label{usr.bin.r}

The \RAppArmor package allows us to dynamically load an \AppArmor profile from
within an \R session. This gives a great deal of flexibility. However, it is
also possible to use \AppArmor without the \RAppArmor package, by setting a
single profile to be loaded with any running \R process.

To do so, the RAppArmor package ships with a profile named \texttt{usr.bin.r}.
At the installation of the package, this file is copied to
\texttt{/etc/apparmor.d/}. This file is basically a copy of the \texttt{r-user}
profile in appendix \ref{r-user}, however with a small change: where
\texttt{r-user} defines a named profile with
\begin{verbatim}
  profile r-user {
    ...
  }
\end{verbatim}
the \texttt{usr.bin.r} file defines a profile specific to a filepath:
\begin{verbatim}
  /usr/bin/R {
    ...
   }
\end{verbatim}

When using the latter syntax, the profile is automatically associated every
time the file \texttt{/usr/bin/R} is executed (which is the script that runs
when \R is started from the shell). This way we can set some default security
restrictions for our daily work. Profiles tied to a specific program can be
activated only by the administrator using:
\begin{verbatim}
  sudo aa-enforce usr.bin.r
\end{verbatim}
This will enforce the security restrictions on every new \R process that is
started. To stop enforcing the restrictions, the administrator can run:
\begin{verbatim}
  sudo aa-disable usr.bin.r
\end{verbatim}
After disabling the profile, the \R program can be started without any
restrictions.
Note that the \texttt{usr.bin.r} profile does \textbf{not} grant permission to
change profiles. Hence, once the \texttt{usr.bin.r} profile is in enforce mode,
we cannot use the \texttt{eval.secure} or \texttt{aa\_change\_profile} functions
from the \RAppArmor package to change into a different profile, as this
would be a security hole:

\begin{CodeChunk}
\begin{CodeInput}
R> library(RAppArmor)
AppArmor LSM is enabled.
Current profile: /usr/bin/R (enforce mode)
R> aa_change_profile("r-user")
Switching profiles...
Getting task confinement information...
Error in aa_change_profile("r-user") : 
  Failed to change profile from: /usr/bin/R to: r-user.
  Note that this is only allowed if the current profile has a
  directive "change_profile -> r-user".
\end{CodeInput}
\end{CodeChunk}

\subsection{Learning using complain mode}

Finally \AppArmor allows the administrator to set profiles in \emph{complain
mode}, which is also called \emph{learning mode}.
\begin{verbatim}
  sudo aa-complain usr.bin.r
\end{verbatim}
This is useful for developing new profiles. When a profile is set in complain
mode, security restrictions are not actually enforced; instead all violations
of the security policy are logged to the \texttt{syslog} and \texttt{kern.log}
files. This is a powerful way of creating new profiles: a program can be set in
complain mode during regular use, and afterwards the log files can be used to
study violations of the current policy. From these violations we can determine
which permissions need to be added to the profile to make the program work
under normal behavior. \AppArmor even ships with a utility named
\texttt{aa-logprof} which can help the administrator by parsing these log files and
suggesting new rules to be added to the profile. This is a nice way of
debugging a profile, and figure out which permissions exactly a program
requires to do its work.

\section[Profiling R: Defining security policies]{Profiling \R: Defining
security policies}

The ``hard'' part of the problem is actually profiling \R. With
profiling we mean defining the policies: which files and directories should
\R be allowed to read and write to? Which external programs is it
allowed to execute? Which libraries or shared modules it allowed to load, etc.
We want to minimize ways in which the process could potentially damage the
system, but we don't want to be overly restrictive either: preferebly, users
should be able to do anything they normally do in \R. Because \R is
such a complete system with a big codebase and a wide range of functionality,
the base system actually already requires quite a lot of access to the file
system.

As often, there is no ``one size fits all'' solution. Depending on which
functionality is needed for an application we might want to grant or deny
certain privileges. We might even want to execute some parts of a process with
tighter privileges than other parts. For example, within a web service, the
service process should be able to write to system log files, which should not be
writable by custom code from a user. We might also want to be more strict on
some users than others, e.g.\ allow all users to run code, but only allow
privileged users to install a new package.

\subsection[AppArmor policy configuration syntax]{\AppArmor policy configuration
syntax}
\label{syntax}

The \emph{AppArmor policy configuration syntax} is used to define the access
control profiles in \AppArmor. Other mandatory access control systems
might implement different functionality and require other syntax, but in the end
they address mostly similar issues. \AppArmor is quite advanced and provides
access control over many of the features and resources found in the \Linux
kernel, e.g.\ file access, network rules, \Linux capability modes, mounting
rules, etc. All of these can be useful, but most of them are very application
specific. Furthermore, the policy syntax has some meta functionality that
allows for defining \emph{subprofiles}, and \emph{includes}.

The most important form of access control which will be the focus of the
remaining of the section are \emph{file permission access modes}. Once \AppArmor
is enforcing mandatory access control, a process can only access files and
directories on the system for which it has explicitly been granted access in
its security profile. Because in \Linux almost everything is a file
(even sockets, devices, etc) this gives a great deal of control. \AppArmor
defines a number of access modes on files and directories, of which the most
important ones are:

\begin{itemize}
  \item[] \texttt{r} -- read file or directory.
  \item[] \texttt{w} -- write to file or directory.
  \item[] \texttt{m} -- load file in memory.
  \item[] \texttt{px} -- discrete profile execute of executable file.
  \item[] \texttt{cs} -- transition to subprofile for executing a file.
  \item[] \texttt{ix} -- inherit current profile for executing a file.
  \item[] \texttt{ux} -- unconfined execution of executable file (dangerous).
\end{itemize}

Using this syntax we will present some example profiles for \R. Because the
profiles are defined using absolute paths of system files, we will assume the
standard file layout for Debian and Ubuntu systems. This includes files that
are part of \texttt{r-base} and other packages that are used by \R, e.g.
\texttt{texlive}, \texttt{libxml2}, \texttt{bash}, \texttt{libpango},
\texttt{libcairo}, etc.

\subsection[Profile: r-base]{Profile: \texttt{r-base}}
\label{r-base-intro}

Appendix \ref{r-base} contains a profile that we have named \texttt{r-base}.
It is a fairly basic and general profile. It grants read/load access to all
files in common shared system directories, e.g.\ \texttt{/usr/lib,
/usr/local/lib, /usr/share}, etc. However, the default profile only grants
write access inside \texttt{/tmp}, not in e.g.\ the home directory. Furthermore,
\R is allowed to execute any of the shell commands in \texttt{/bin}
or \texttt{/usr/bin} for which the program will inherit the current
restrictions.

\begin{CodeChunk}
\begin{CodeInput}
R> library("RAppArmor")
R> aa_change_profile("r-base")
Switching profiles...

R> list.files("/")
character(0)
R> list.files("~")
character(0)
R> file.create("~/test")
[1] FALSE
R> list.files("/tmp")
character(0)
R> install.packages("wordcloud")
Error opening file for reading: Permission denied

R> library("ggplot2")
R> setwd(tempdir())
R> pdf("test.pdf")
R> qplot(speed, dist, data=cars)
R> dev.off()
null device
          1
R> list.files()
[1] "downloaded_packages"
[2] "libloc_107_669a3e12.rds"
[3] "libloc_118_46fd5f8e.rds"
[4] "libloc_128_97f33314.rds"
[5] "pdf6d1117f7d683"
[6] "repos_http%3a%2f%2fcran.stat.ucla.edu%2fsrc%2fcontrib.rds"
[7] "test.pdf"
R> file.remove("test.pdf")
[1] TRUE
\end{CodeInput}
\end{CodeChunk}

The \texttt{r-base} profile effectively protects \R from most malicious
activity, while still allowing access to all of the libraries, fonts, icons,
and programs that it might need. One thing to note is that the profile does not
allow listing of the contents of \texttt{/tmp}, but it does allow full
\texttt{rw} access on any of its subdirectories. This is to prevent one process
from reading/writing files in the temp directory of another active \R process
(given that it cannot discover the name of the other temp directory).

The \texttt{r-base} profile is a quite liberal and general purpose profile.
When using \AppArmor in a more specific application, it is recommended to make
the profile a bit more restrictive by specifying exactly \emph{which} of the
packages, shell commands and system libraries should be accessible by the
application. That could prevent potential problems when vulnerabilities are
discovered in some of the standard libraries.

\subsection[Profile: r-compile]{Profile: \texttt{r-compile}}

The \texttt{r-base} profile does not allow access to the compiler, nor does it
allow for loading (\texttt{m}) or execution (\texttt{ix}) of files in places
where it can also write. If we want user to be able to compile e.g. \Cpp code,
the policy needs grant access to the compiler. Assuming \texttt{GCC} is
installed, the following lines can be added to the profile:

\begin{verbatim}
/usr/include/** r,
/usr/lib/gcc/** rix,
/tmp/** rmw,
\end{verbatim}

Note especially the last line. The combination of \texttt{w} and \texttt{m}
access modes allows \R to load a shared object into memory from after
installing it in a temporary directory. This does not come without a cost:
compiled code can potentially contain malicious code or even exploits that can
do harm when loaded into memory. If this privilege is not needed, it is
generally recommended to only allow \texttt{m} and {ix} access modes on files
that have been installed by the system administrator. The new profile including
these rules ships with the package as \texttt{r-compile} and is also printed in
appendix \ref{r-compile}.

After adding the lines above and reloading the profile, it should be possible to
compile a package that contains \Cpp code and install it to somewhere
in \texttt{/tmp}:

\begin{CodeChunk}
\begin{CodeInput}
R> eval.secure(install.packages("wordcloud", lib=tempdir()), profile="r-compile")
trying URL 'http://cran.stat.ucla.edu/src/contrib/wordcloud_2.0.tar.gz'
downloaded 36 Kb

* installing *source* package 'wordcloud' ...
** package 'wordcloud' successfully unpacked and MD5 sums checked
** libs
g++ -I/usr/share/R/include -DNDEBUG -I"/usr/local/lib/R/site-library/Rcpp/include"
   -fpic  -O3 -pipe  -g  -c layout.cpp -o layout.o
g++ -shared -o wordcloud.so layout.o -L/usr/local/lib/R/site-library/Rcpp/lib
   -lRcpp -Wl,-rpath,/usr/local/lib/R/site-library/Rcpp/lib -L/usr/lib/R/lib -lR
installing to /tmp/RtmpFCM6WS/wordcloud/libs
** R
** data
** preparing package for lazy loading
** help
*** installing help indices
** building package indices
** testing if installed package can be loaded

* DONE (wordcloud)

The downloaded source packages are in
	'/tmp/RtmpFCM6WS/downloaded_packages'
\end{CodeInput}
\end{CodeChunk}

\subsection[Profile: r-user]{Profile: \texttt{r-user}}

Appendix \ref{r-user} defines a profile named \texttt{r-user}. This profile is
designed to be a nice balance between security and freedom for day to day use
of R. It extends the \texttt{r-compile} profile with some additional privileges
in the user's home directory. The variable \texttt{@\{HOME\}} is defined in the
\texttt{/etc/apparmor.d/tunables/global} include that ships with \AppArmor and
matches the location of the user home directory, i.e.\ \texttt{/home/jeroen/}.
If a directory named \texttt{R} exists inside the home directory 
(e.g \texttt{/home/jeroen/R/}), \R has both read and write permissions here.
Furthermore, \R can load and execute files in the directories
\texttt{i686-pc-linux-gnu-library} and \texttt{x86\_64-pc-linux-gnu-library}
inside of this directory. These are the standard locations where \R installs a
user's personal package library.

With the \texttt{r-user} profile, we can do most of our day to day work,
including installing and loading new packages in our personal library, while
still being protected against most malicious activities. The \texttt{r-user}
profile is also the basis of the default \texttt{usr.bin.r} profile mentioned
in section \ref{usr.bin.r}.

\subsection{Installing packages}

An additional privilege that might be needed in some situations is the option
to install packages to the system's global library, which is readable by all
users. In order to allow this, a profile needs to include write access to the
\texttt{site-library} directory:

\begin{verbatim}
 /usr/local/lib/R/site-library/ rw,
 /usr/local/lib/R/site-library/** rwm,
\end{verbatim}

With this rule, the policy will allow for installing \R packages to the global
site library. However, note that \AppArmor does not replace, but
\emph{supplements} the standard access control system. Hence if a user does not
have permission to write into this directory (either by standard Unix access
controls or by running with superuser privileges), it will still not be able to
install packages in the global site library, even though the \AppArmor profile
does grant this permission.

\section{Concluding remarks}

In this paper the reader was introduced to some potential security issues
related to the use of the \R software. We hope to have raised awareness that
security is an increasingly important concern for the \R user, but also that
addressing this issue could open doors to new applications of the software. The
\RAppArmor package was introduced as an example that demonstrates how some
security issues could be addressed using facilities from the operating system,
in this case \Linux. This implementation provides a starting point for creating
a sandbox in \R, but as was emphasized throughout the paper, it is still up to
the administrator to actually design security policies that are appropriate for
a certain application or system.

Our package uses the \AppArmor software from the \Linux kernel, which works for
us, but this is just one of the available options. \Linux has two other
mandatory access control systems that are worth exploring: \texttt{TOMOYO} and
\texttt{SELinux}. Especially the latter is known to be very sophisticated, but
also extremely hard to set up. Other technology that might be interesting is
provided by \texttt{Linux CGroups}. Using \texttt{CGroups}, control of
allocation and security is managed by hierarchical process groups. The more
recent \texttt{LXC} (Linux Containers) build on \texttt{CGroups} to provide
virtual environments which have their own process and network space. A
completely different direction is suggested by \texttt{renjin} \citep{renjin},
a \texttt{JVM}-based interpreter for the R Language. If \R code can be executed
though the \texttt{JVM}, we might be able to use some tools from the \Java
community to address similar issues. Finally the \texttt{TrustedBSD} project
provides advanced security features which could provide a foundation for
sandboxing \R on \texttt{BSD} systems.

However, regardless of the tools that are used, security always comes down to
the trade off between \emph{use} and \emph{abuse}. This has a major human
aspect to it, and is a learning process in itself. A balance has to be found
between providing enough freedom to use facilities as desired, yet minimize
opportunities for undesired activity. Apart from technical parameters, a good
balance also depends on factors like what exactly constitutes undesired
behavior and the relation between users and provider. For example a process
using 20 parallel cores might be considered abusive by some administrators, but
might actually be regular use for a MCMC simulation server. Security policies
are not unlike legal policies in the sense that they won't always immediately
work out as intended, and need to evolve over time as part of an iterative
process. It might not be until an application is put in production that users
start complaining about their favorite package not working, or that we find the
system being abused in a way that was hard to foresee. We hope that our
research will contribute to this process and help take a step in the direction
of a safer \R.

\section{Acknowledgments}

We owe gratitude to several people who have been specifically helpful in the
course of this research. Things would not have been possible without their 
valuable criticism, support and feedback. Among others these include Dirk
Eddelbuettel and Michael Rutter for providing excellent packages for the Debian
and Ubuntu distributions on which we largely build our implementation. Daróczi
Gergely and Aleksandar Blagotić for always being ``early adopters'' (guinea
pigs) and putting things to the test. And finally John Johansen, Seth Arnold and
Steve Beattie have been very helpful (and patient) by providing support and
feedback through the \texttt{apparmor} mailing lists.

\begin{appendices}
\section{Example profiles}

This appendix prints some of the example profiles that ship with the \RAppArmor
package. To load them in \AppArmor, an ascii file with these rules needs to be
copied to the \texttt{/etc/apparmor.d/} directory. After adding new profiles to
this directory they can be loaded in the kernel by running \texttt{sudo service
apparmor restart}. The \texttt{r-cran-rapparmor} package that can be build on
Debian and Ubuntu does this automatically during installation. Once profiles
have been loaded in the kernel, any user can apply them to an \R session using
either the \texttt{aa\_change\_profile} or \texttt{eval.secure} function from
the \texttt{RAppArmor} package.

\subsection[Profile: r-base]{Profile: \texttt{r-base}}
\label{r-base}

\begin{verbatim}
#include <tunables/global>
profile r-base {
        #include <abstractions/base>
        #include <abstractions/nameservice>

        /bin/* rix,
        /etc/R/ r,
        /etc/R/* r,
        /etc/fonts/** mr,
        /etc/xml/* r,
        /tmp/** rw,
        /usr/bin/* rix,
        /usr/lib/R/bin/* rix,
        /usr/lib{,32,64}/** mr,
        /usr/lib{,32,64}/R/bin/exec/R rix,
        /usr/local/lib/R/** mr,
        /usr/local/share/** mr,
        /usr/share/** mr,
}
\end{verbatim}

\subsection[Profile: r-compile]{Profile: \texttt{r-compile}}
\label{r-compile}

\begin{verbatim}
#include <tunables/global>
profile r-compile {
        #include <abstractions/base>
        #include <abstractions/nameservice>

        /bin/* rix,
        /etc/R/ r,
        /etc/R/* r,
        /etc/fonts/** mr,
        /etc/xml/* r,
        /tmp/** rmw,
        /usr/bin/* rix,
        /usr/include/** r,
        /usr/lib/gcc/** rix,		
        /usr/lib/R/bin/* rix,
        /usr/lib{,32,64}/** mr,
        /usr/lib{,32,64}/R/bin/exec/R rix,
        /usr/local/lib/R/** mr,
        /usr/local/share/** mr,
        /usr/share/** mr,
}
\end{verbatim}

\subsection[Profile: r-user]{Profile: \texttt{r-user}}
\label{r-user}

\begin{verbatim}
#include <tunables/global>
profile r-user {
        #include <abstractions/base>
        #include <abstractions/nameservice>
	
        capability kill,
        capability net_bind_service,
        capability sys_tty_config,
	
        @{HOME}/ r,
        @{HOME}/R/ r,
        @{HOME}/R/** rw,
        @{HOME}/R/{i686,x86_64}-pc-linux-gnu-library/** mrwix,
        /bin/* rix,
        /etc/R/ r,
        /etc/R/* r,
        /etc/fonts/** mr,
        /etc/xml/* r,
        /tmp/** mrwix,
        /usr/bin/* rix,
        /usr/include/** r,
        /usr/lib/gcc/** rix,		
        /usr/lib/R/bin/* rix,
        /usr/lib{,32,64}/** mr,
        /usr/lib{,32,64}/R/bin/exec/R rix,
        /usr/local/lib/R/** mr,
        /usr/local/share/** mr,
        /usr/share/** mr,
}
\end{verbatim}

\section{Security unit tests}

This appendix prints a number of unit tests that contain malicious code and
which should be prevented by any sandboxing tool.

\subsection{Access system files}

Usually \R has no business in the system logs, and these are not included in
the profiles. The codechunk below attempts to read the syslog file.
\begin{CodeChunk}
\begin{CodeInput}
readSyslog <- function(){
	readLines('/var/log/syslog')
}
\end{CodeInput}
\end{CodeChunk}
When executing this with the r-user profile, access to this file is denied,
resulting in an error:
\begin{CodeChunk}
\begin{CodeInput}
R> eval.secure(readSyslog(), profile='r-user')
Switching profiles...
Error in file(con, "r") : cannot open the connection
\end{CodeInput}
\end{CodeChunk}

\subsection{Access personal files}
\label{creditcard}

Access to system files can to some extend by prevented by running processes as
non privileged users. But it is easy to forget that also the user's personal
files can contain senstive information. Below a simple function that scans the
\texttt{Documents} directory of the current user for files containing credit
card numbers.

\begin{CodeChunk}
\begin{CodeInput}
findCreditCards <- function(){
  pattern <- "([0-9]{4}[- ]){3}[0-9]{4}"
  for (filename in list.files("~/Documents", full.names=TRUE, recursive=TRUE)){
    if(file.info(filename)$size > 1e6) next
    doc <- readLines(filename)
    results <- gregexpr(pattern, doc)
    output <- unlist(regmatches(doc, results))
    if(length(output) > 0){
      cat(paste(filename, ":", output, collapse="\n"), "\n")
    }
  }
}
\end{CodeInput}
\end{CodeChunk}

This example prints the credit card numbers to the console, but it would be
just as easy to post them to a server on the internet. For this reason the
\texttt{r-user} profile denies access to the user's home dir, except for the
\texttt{{\raise.17ex\hbox{$\scriptstyle\sim$}}/R} directory.

\subsection{Limiting memory}

When a system or service is used by many users at the same time, it is
important that we cap the memory that can be used by a single process. The
following function generates a large matrix:

\begin{CodeChunk}
\begin{CodeInput}
memtest <- function(){
	A <- matrix(rnorm(1e7), 1e4)
}
\end{CodeInput}
\end{CodeChunk}

When \R tries to allocate more memory than allowed, it will throw an error:

\begin{CodeChunk}
\begin{CodeInput}
R> A <- eval.secure(memtest(), RLIMIT_AS = 1000*1024*1024)
R> rm(A)
R> gc()
         used (Mb) gc trigger  (Mb) max used  (Mb)
Ncells 193074 10.4     407500  21.8   350000  18.7
Vcells 299822  2.3   17248096 131.6 20301001 154.9

R> A <- eval.secure(memtest(), RLIMIT_AS = 100*1024*1024)
Error: cannot allocate vector of size 76.3 Mb
\end{CodeInput}
\end{CodeChunk}

\subsection{Limiting CPU time}
\label{cputime}

Suppose we are hosting a web service and we want to kill jobs that do not
finish within 5 seconds. Below is a snippet that will take much more than 5
seconds to complete on most machines. Note that because \R calling out to
\texttt{C} code, it will not be possible to terminate this function prematurely
using R's \texttt{setTimeLimit} or even using \texttt{CTRL+C} in an interactive
console. If this would happen inside of a bigger system, the entire service
might become unresponsive.

\begin{CodeChunk}
\begin{CodeInput}
cputest <- function(){
  A <- matrix(rnorm(1e7), 1e3)
  B <- svd(A)
}
\end{CodeInput}
\end{CodeChunk}
In RAppArmor we have actually two different options to deal with this. The first
one is setting the \texttt{RLIMIT\_CPU} value. This will cause the kernel to
kill the process after 5 seconds:
\begin{CodeChunk}
\begin{CodeInput}
R> system.time(x <- eval.secure(cputest(), RLIMIT_CPU=5))
   user  system elapsed 
  0.004   0.000   5.110 
R> print(x)
NULL
\end{CodeInput}
\end{CodeChunk}
However, this is actually a bit of a harsh measure: because the kernel
automatically terminates the process after 5 seconds we have no control over
what should happen when this happens, nor can we throw an informative error.
Setting \texttt{RLIMIT\_CPU} is a bit like starting a job with a
self-destruction timer. A more elegant solution is to terminate the process
from \R using the \texttt{timeout} argument from the \texttt{eval.secure}
function. Because the actual job is processed in a fork, the parent process
stays responsive, and is used to kill the child process.
\begin{CodeChunk}
\begin{CodeInput}
R> system.time(x <- eval.secure(cputest(), timeout=5))
Error: R call did not return within 5 seconds. Terminating process.
Timing stopped at: 4.748 0.26 5.007
\end{CodeInput}
\end{CodeChunk}

One could even consider a Double Dutch solution by setting both \texttt{timeout}
and a slightly higher value for \texttt{RLIMIT\_CPU}, so that if all else fails,
the kernel will end up killing the process and its children.

\subsection{Fork bomb}

A fork bomb is a process that spawns many child processes, which often results
in the operating system getting stuck to a point where it has to be rebooted.
Performing a fork bomb in \R is quite easy and requires no special privileges:
\begin{CodeChunk}
\begin{CodeInput}
forkbomb <- function(){
  repeat{
    parallel::mcparallel(forkbomb())
  }
}
\end{CodeInput}
\end{CodeChunk}
Do not call this function outside sandbox, because it will make the machine
unresponsive. However, inside our sandbox we can use the \texttt{RLIMIT\_NPROC}
to limit the number of processes the user is allowed to own:
\begin{CodeChunk}
\begin{CodeInput}
R> eval.secure(forkbomb(), RLIMIT_NPROC = 20)
RLIMIT_NPROC:
Previous limits: soft=39048; hard=39048
Current limits: soft=20; hard=20
Error in mcfork() :
  unable to fork, possible reason: Resource temporarily unavailable
\end{CodeInput}
\end{CodeChunk}
Note that the process count is based on the \Linux user. Hence if the same
\Linux user already has a number of other processes, which is usually the case
for non-system users, the cap has to be higher than this number. Also note that
in some \Linux configurations, the root user is exempted from the
\texttt{RLIMIT\_NPROC} limit.

Different processes owned by a single user can enforce different \texttt{NPROC}
limits, however in the actual process count all active processes from the
current user are taken into account. Therefore it might make sense to create a
separate Linux system user that is only used to process \R jobs. That way
\texttt{RLIMIT\_NPROC} actually corresponds to the number of concurrent \R
processes. The \texttt{eval.secure} arguments \texttt{uid} and \texttt{gid}
can be used to switch Linux users before evaluating the call. E.g to add a
system user in \Linux, run:
\begin{CodeChunk}
\begin{CodeInput}
sudo useradd testuser --system -U -d/tmp -c"RAppArmor Test User"
\end{CodeInput}
\end{CodeChunk}
If the main \R process has superuser privileges, incoming call can be
evaluated as follows:
\begin{CodeChunk}
\begin{CodeInput}
eval.secure(run_job(), uid="testuser", RLIMIT_NPROC=10, timeout=60)
\end{CodeInput}
\end{CodeChunk}

\end{appendices}

\bibliography{v55i07}

\end{document}